\documentstyle[12pt,aaspp]{article}

\begin{document}

\title{ROSAT Observations of Compact Groups of Galaxies}

\author{Rachel A. Pildis and Joel N. Bregman}
\affil{Department of Astronomy, University of Michigan, Ann Arbor,
	Michigan 48109-1090\\ I: pildis@astro.lsa.umich.edu,
	jbregman@astro.lsa.umich.edu}

\author{August E. Evrard}
\affil{Department of Physics, University of Michigan, Ann Arbor,
	Michigan 48109-1120\\ I: evrard@umich.edu}

\begin{abstract}

We have systematically analyzed a sample of 13 new and archival ROSAT PSPC
observations of compact groups of galaxies: 12 Hickson Compact Groups plus
the NCG 2300 group.  We find that approximately two-thirds of the groups
have extended X-ray emission and, in four of these, the emission is resolved
into diffuse emission from gas at a temperature of $k$T $\sim 1$ keV in the
group potential.  All but one of the groups with extended emission have a
spiral fraction of less than 50\%.  The baryon fraction of groups with
diffuse emission is 5--19\%, similar to the values in clusters of galaxies.
However, with a single exception (HCG 62), the gas-to-stellar-mass ratio in
our groups has a median value near 5\%, somewhat greater than the values for
individual early-type galaxies and two orders of magnitude lower than in
clusters of galaxies.  The X-ray luminosities of individual group galaxies
are comparable to those of similar field galaxies, although the L$_X$-L$_B$
relation for early-type galaxies may be flatter in compact groups than in
the field.

\end{abstract}


\section{Introduction}

	Poor groups of galaxies have been studied much less than have rich
clusters, even though they are the most common type of galaxy
grouping and thus are more likely to be representative of the universe as a
whole than relatively rare clusters.  In large part, this is because groups
are much more difficult to identify:  they have very few galaxies and tend
to have small overdensities relative to the background.  Even those groups
selected by ``objective'' searches through redshift space such as the CfA
groups (Huchra \& Geller 1982) may not all be gravitationally bound (Moore,
Frenk, \& White 1993).  These concerns lead one to search for groups with
few galaxies that nonetheless have relatively large overdensities for easy
identification; such groups do exist---compact groups of galaxies.

	Compact groups contain some of the densest concentrations of
galaxies known, with galaxy densities comparable to or greater than even
those of the cores of rich clusters.  These groups typically contain fewer
than 10 members, and the galaxy-galaxy separations are on the order of a
galaxy diameter.  The high densities would imply that these are ideal
laboratories for studying galaxy interactions and merging.

	One well-studied sample of compact groups is the Hickson Compact
Groups (HCG's) as defined by Hickson (1982).  Spectroscopy and imaging have
shown that most of the groups consist of galaxies with concordant redshifts
(Hickson et al.~1992) and that many of these galaxies display signs of
interaction such as tidal tails and distorted spiral arms (Mendes de
Oliveira \& Hickson 1994; Rubin, Hunter, \& Ford 1991).  However, studies of
early-type galaxies in HCG's find them to be less boxy than field
ellipticals (Fasano \& Bettoni 1994, Zepf \& Whitmore 1993), which might
imply that these galaxies have undergone fewer interactions than have field
ellipticals.

	The question of the interaction history of HCG's is not answered by
studies at radio and infrared frequencies.  Several investigations into the
neutral hydrogen distribution in various HCG's (Shostak, Sullivan, \& Allen
1984; Williams \& van Gorkom 1988; Williams, McMahon, \& van Gorkom 1991)
have shown that some of the more compact groups have HI clouds that
encompass entire groups, while the less compact groups have HI confined to
the (spiral) galaxies.  The galaxies in these latter groups contain only
half the neutral hydrogen that similar galaxies in loose groups have
(Williams \& Rood 1987).  Hickson et al.~(1989) found that the IRAS FIR
emission in compact groups was enhanced, but this was disputed by Sulentic
\& de Mello Raba\c{c}a (1993).  Moles et al.~(1994) considered
25\micron\ emission in addition to FIR emission, and found no enhancement
over the levels seen in isolated pairs of galaxies.

	One can determine that a group is gravitationally bound by detecting
hot gas in the group potential well via X-ray observations.  Few such
observations of compact groups have been made, but this situation is rapidly
changing.  Bahcall, Harris, \& Rood (1984) observed four HCG's (and an
additional Arp group) with the Einstein X-ray telescope and found extended
emission in HCG 92 (Stephan's Quintet) and Arp 330.  The sensitivity to
emission at temperatures near 1 keV of the ROSAT X-ray telescope has made
possible a great number of observations of compact groups of galaxies, two
of which have already been reported upon in the literature.  HCG 62 was
found to be a bright source of diffuse X-rays by Ponman \& Bertram (1993),
and the NGC 2300 group of galaxies was found to have extended X-ray emission
by Mulchaey et al.~(1993).  The latter paper claimed that the NGC 2300 group
has a baryon fraction of only 4\%\ (although this increases to at least
6\%\ using their note added in proof), which would indicate that these systems
are considerably different than rich clusters (which have baryon fractions
of 10--30\%) and very similar to some theoretical predictions for the
baryonic content of the universe as a whole (e.g., White et al.~1993).

	In order to determine whether these two groups are typical of compact
groups, we assembled a large sample of X-ray observations of HCG's.  In
addition to our PSPC observations of four HCG's (selected to have a large
enough angular size to be resolved by the PSPC, and far enough north for
follow-up optical observations at MDM Observatory on Kitt Peak), we obtained
all the archived ROSAT PSPC observations of HCG's available as of April 1994
(including those of HCG 62, and the NGC 2300 group).  This resulted in a
sample of 13 groups, which we then reduced in a uniform manner.  The PSPC
observations allowed us to find the fraction of groups with extended
emission, and then determine the mass of hot gas and the total gravitating
mass of those systems.  These masses were compared to the baryonic mass in
the galaxies of each group in order to determine gas-to-stellar-mass ratios
and baryonic fractions.  This study is an improvement over previous work
since the flattening properties of the PSPC are better understood and the
background subtraction was investigated thoroughly.  In addition, the
reduction and analysis was uniformly applied to all the objects, so they
could be compared to one another.

\section{Observations and Data Reduction}

	Four HCG's were observed with the PSPC as part of our observing
program (HCG 2, 10, 68, and 93), while seven other HCG's and the
NCG 2300 group had PSPC observations obtained from the ROSAT archives.
In addition, HCG 94 can be seen in the observation of HCG 93, although
it is over 30\arcmin\ off axis.  A summary of the properties of the groups
is in Table 1 (throughout this paper, H$_0$=50 km s$^{-1}$ Mpc$^{-1}$ and
$q_0$=0 are assumed).

	All observations were reduced in the same manner.  First, we
examined the background level as a function of time by binning the counts in
a $\sim$100 square arcminute region containing no apparent X-ray sources in
roughly 100 second intervals.  We examined the value of the signal-to-noise
for a source 10\%\ as bright as the modal background level as we included
time bins with greater and greater background levels.  Typically, the
signal-to-noise would rise steadily as more bins were included, but fall
for those bins at very high background levels.  We then excluded all
time intervals which had background levels above the point where the
signal-to-noise ratio started to decrease.  In all but one case (HCG 92), this
entailed the exclusion of less than 8\% of the total observing time.
Subsequent analysis was performed using these filtered observations.  Table
2 lists the total and net observing times, as well as the galactic neutral
hydrogen column (N$_{\rm H}$) in the direction of each group as determined
by Stark et al.~(1992).

	Each PSPC field was flattened in the appropriate energy band using
the program CAST\_EXP (Snowden et al.~1994), which creates more accurate
flat fields than those provided by the US ROSAT Science Data Center, which
are not energy-band specific.  This program
uses the aspect and event rate information for each observation, along with
the time intervals used and the energy band of interest, to make a
flat field.  These flat fields were used to make flat images for
analyzing the extent of any X-ray emission, as well as to determine
normalization factors for spectral analysis.  The flat images excluded any
energy channels that would be absorbed heavily by the neutral hydrogen
column in our Galaxy (e.g., $k$T $<$ 0.5 keV for high columns).

	In order to test the accuracy of our field flattening, we flattened
three fields from the ROSAT archives (rp700112, rp700117, and rp700375) that
had relatively few sources, none of them expected to be extended.  These
three fields intentionally span a range of neutral hydrogen columns (N$_{\rm
H}$ = 4.07, 1.20, and 2.10 $\times  10^{20}$ cm$^{-2}$, respectively) and
exposure times (19.8, 22.7, and 13.4 ksec) in order to see what effect
these parameters had on the flattening.  None had more than 1 ksec of observing
time excluded due to high background levels.

	A plot of the background level as a function of distance from
field center for the two high N$_{\rm H}$ fields (rp700112 and rp700375)
is shown in Fig.~1.  The data points are averages of annular bins 40\arcsec\
wide.  The background level becomes less well determined at the radius of
the support ring (r $\sim$ 20\arcmin ) and beyond, as well as at radii less
than 7\arcmin , where the targets of the observation may not be adequately
excluded.  Further analysis shows
that the fluctuations of the background for 7\arcmin\ $<$ r $<$ 20\arcmin\
are approximately 1.4 times greater than that expected from photon statistics
alone.  It is unclear whether this is due to unsubtracted point sources,
fluctuations in N$_{\rm H}$, imperfections in the flattening process, or
a combination of all these.  Whatever the cause, this larger error should
be applied when analyzing PSPC data, especially when searching for
extended emission that is only a fraction of the background level.

	The low N$_{\rm H}$ field, rp700117, could not be flattened to the
same degree of accuracy as the other two fields.  Not only was it more
difficult to create a good flat when nearly all the energy channels were
included, but this field exhibited an increase in surface brightness with
radius, regardless of the energy band examined (Fig.~2); especially for
r~$>$~20\arcmin .  Even when channels with $k$T $<$ 0.5 keV were excluded
(heavy line), the background counts rise with radius even within the support
ring.  It is not clear whether this is an artifact of the flattening process
or some systematics of the field or observation (or some combination of
these).  This demonstrates that processing PSPC images is not guaranteed to
produce perfectly flat fields, with the greatest difficulties at r $\geq$
20\arcmin .  Again, extreme caution must be exercised when analyzing low
surface brightness X-ray features---it can be highly difficult to
disentangle such features from telescope or reduction systematics.

\section{Analysis and Discussion}

	All of the spatial and the initial spectral reduction was done using
the PROS package within IRAF, while the remainder of the spectral reduction
used XSPEC.  Our procedure was as follows:  after the initial
reduction outlined in \S 2, a flattened image of the central square degree
of the PSPC field was created, using only energy channels largely unabsorbed
by neutral hydrogen in the Galaxy.  The flat field from CAST\_EXP was
normalized so that the maximum pixel value is 1.  A smoothed image (smoothed
with a gaussian with $\sigma$=16\arcsec) was then created, and
viewed with an overlay of the optical galaxy positions.  Contour maps of the
X-ray emission of the groups with detected extended emission are in Fig.~3.
Each smoothed image was visually examined to see if any of the galaxies in
the group have X-ray emission and/or if there is any extended X-ray emission
associated with the group.  If there was extended emission, we used annular
binning on the unsmoothed data to determine its extent.

	All galaxies identified in the flattened and smoothed image were
spectrally analyzed, as was all extended emission.  A background-subtracted,
vignetting-corrected spectrum of each detection was analyzed in XSPEC, where
we fit an absorbed Raymond-Smith model to the data, fixing the neutral
hydrogen column to the galactic column in the direction of the group.  Since
the PSPC does not allow one to fit abundances very well, we fixed abundances
of diffuse gas at one-half solar and galaxies at solar abundance.  If the
source was too faint to fit a temperature, we fixed it to be $k$T=0.5 keV
(elliptical and S0 galaxies) or 1.0 keV (spiral galaxies and diffuse
emission). The fitted model then was used to find the rest frame luminosity
of the source in the XSPEC PSPC passband (0.07--3.0 keV).  The diffuse
emission will be discussed in \S 3.3.

	For galaxies in the PSPC field but not detected at the $\gtrsim
3\sigma$ level, we determined upper limits.  First, we found the net number
of photons in a circular aperture of radius equal to the major axis diameter
of the galaxy, as given in Hickson (1993).  Then we obtained a luminosity
(if the net counts were 2$\sigma$ or greater above the background---such
galaxies could be seen in the smoothed image) or a
3$\sigma$ upper limit on the luminosity for each galaxy by scaling the
luminosity of a well-detected galaxy of the same type in the same group by
the ratio of the net counts of the two galaxies.  If there was not a galaxy
of the same type (E/S0 vs.~spiral) in the group, we forced a fit of the
appropriate temperature on a well-detected galaxy in the group, and used the
result to obtain a scaled luminosity.  This procedure was not carried out
for galaxies whose emission could not be separated from nearby bright
emission (e.g., the galaxies of HCG 4).  Table 3 summarizes the X-ray and
optical properties of all PSPC-detected compact group galaxies.  We use the
galaxy designations of Hickson (1982, 1993) throughout this paper.

\subsection{Individual Groups---Comments}

	Eight of the 13 groups we analyzed have detectable extended X-ray
emission (see Table 4).  Unfortunately, several of the groups are not
completely resolved with the PSPC (which has a FWHM of 25\arcsec)---it was
impossible to separate diffuse emission from emission from the galaxies.
Also, some of the same groups had observing times that were too short to
obtain high signal-to-noise ratio detections of the extended emission.  Only 4
groups had resolved detections of diffuse X-ray gas in the potential well of
the group.  The observations of each of the 13 groups are discussed more
thoroughly below.

\paragraph{HCG 2}

	All three of the galaxies in this group with accordant redshifts
are detected in X-rays, but no extended emission is seen.

\paragraph{HCG 4}

	A bright power-law source in galaxy {\it a} of this group overwhelms
any possible signal from other sources in the group.  This source has a
power-law index $\alpha = 2.08 \pm 0.04$ (where the emission depends on the
energy as E$^{-\alpha}$) and a luminosity in the PSPC passband of $1.1
\times 10^{43}$ erg s$^{-1}$.  Galaxy {\it a} was found to contain a
possible AGN in the Montreal Blue Galaxy Survey (Coziol et al.~1993), and
this observation appears to confirm the presence of an active nucleus in
that galaxy.

\paragraph{HCG 10}

	Two of the four galaxies in this group ({\it a} and {\it b})
are well-detected by the PSPC, but no extended emission is seen.

\paragraph{HCG 12}

	This quintet of galaxies is not resolved with the PSPC.  The
emission is extended, but the 25\arcsec\ resolution of the instrument
is not sufficient to separate galaxy emission from diffuse emission.

\paragraph{HCG 44}

	This group has three ({\it a}, {\it b}, and {\it c}) of its four
galaxies clearly detected in the PSPC observation, but there is
no extended emission.

\paragraph{HCG 62}

	HCG 62 contains two central, X-ray bright early-type galaxies as
well as very strong diffuse X-ray emission.  It was first analyzed by Ponman
\& Bertram (1993).

\paragraph{HCG 68}

	This group is the only one of our four original PSPC observations
to show diffuse emission.  Four ({\it a--d}) of the five galaxies in
the group are well-detected, although the emission from galaxies {\it a} and
{\it b} overlaps and cannot be resolved separately.

\paragraph{HCG 79}

	This group is better known as Seyfert's Sextet.  Since it is a
very dense, elliptical-rich group, it might well be expected to be a
strong source of X-ray emission; unfortunately, the usable time for this
observation was so short (4.9 ksec) that there were not enough photons
detected to perform any spectral analysis.  In addition, the galaxies in the
group are too close together to be individually resolved with the PSPC
even if the observation time were sufficiently long.

\paragraph{HCG 92}

	The PSPC observation of this group (better known as Stephan's
Quintet) presented some difficulties in reduction.  Not only are
the galaxies too close together to be resolved individually by the
PSPC, but 36\%\ of the original observing time needed to be discarded
because the background level was too high.  Nonetheless, this group
has a strong X-ray detection, though it is unclear whether it is diffuse
emission or unresolved emission from individual galaxies.

\paragraph{HCG 93}

	Only galaxy {\it a} of this group is clearly detected in the PSPC
observation---no extended emission is seen.

\paragraph{HCG 94}

	Serendipitously, this group is seen in the observation of
HCG 93 (it is 30\farcm 5 away from HCG 93).  Because it is so far
off-axis, no determination can be made about the spatial structure
of the X-ray emission, but some spectral analysis can be done (\S 3.3.6).

\paragraph{HCG 97}

	This group has fairly bright extended emission surrounding galaxy
{\it a}, making it difficult to detect or set upper limits on the emission
from any of the individual galaxies of the group.  Even for galaxy {\it a}, we
were forced to assume that all the emission from it was within a radius
of 48\arcsec\ of the center, which may not give an accurate picture of
its X-ray properties.

\paragraph{NGC 2300 group}

	We take this group as consisting of three galaxies:  NGC 2300
(E3, M$_B=-21.3$), NGC 2276 (Sc, M$_B=-21.4$), and IC 455
(S0, M$_B=-19.1$).  This is the same grouping as used by Mulchaey
et al.~(1993), who did the first analysis of the X-ray properties of
this group.


\subsection{Properties of Group Galaxies}

	One question that can be addressed with this sample of compact group
galaxies is whether the apparently very dense environment of these groups
produces systematic differences in the X-ray to optical luminosity
relation.  Such relations have been determined from Einstein observations of
field galaxies and can be easily translated to the ROSAT passband.  Bregman,
Hogg, \& Roberts (1992)[BHR] find a relation for early-type galaxies of L$_X
\sim$ L$_{B}^{2.4}$, and, although the picture for spiral galaxies is less
clear, L$_X$ appears to be linearly correlated with the number of Population
I stars in a galaxy and thus with L$_B$ (Fabbiano \& Trinchieri 1985).

	We have examined the L$_X$-L$_B$ relation for all of the galaxies
detected in the X-ray in our sample that were not contaminated by diffuse
emission or overlapping galaxy emission (questionable cases are noted below
and in Table 3).  Blue luminosities were found using the blue apparent
magnitudes in Hickson (1993), which have been corrected for overlap from
nearby galaxies.  Both the spiral and the early-type galaxies in our sample
appear to follow the field galaxy L$_X$-L$_B$ relations, with a few
exceptions (Fig.~4).  The late-type galaxies follow a linear relation quite
closely; HCG 2b (at log(L$_B$)=10.3, log(L$_X$)=40.2), is somewhat
overluminous in X-rays for its blue luminosity, but it is a Markarian galaxy
and thus may be undergoing some unusual activity.  Among the early-type
galaxies, HCG 97a is the galaxy with the highest L$_X$, and falls about an
order of magnitude above the BHR relation, but this galaxy was difficult to
resolve from the diffuse emission in its group, and its intrinsic emission
may be less than our stated value.  HCG 44b also falls an order of magnitude
above the BHR relation, but has one of the lowest blue luminosities in the
sample; and in the opposite sense, the two elliptical galaxies with the
highest blue luminosities (HCG 10b and HCG 93a) fall over an order of
magnitude below the early-type L$_X$-L$_B$ relation.  None of these three
groups (HCG 10, 44, and 93) has extended emission, so confusion is not an
issue.  HCG 44b and HCG 10b have no known peculiarities and although HCG 93a
is a shell galaxy (Pildis \& Bregman 1995), there is no reason to expect a
relation between optical fine structure and X-ray luminosity.  These results
may indicate that early-type galaxies in compact groups have a flatter
L$_X$-L$_B$ relation than do field galaxies, although this sample is too
small to address this issue definitively.

\subsection{Properties of Groups with Extended Emission}

	For the eight groups in which extended X-ray emission was detected, we
determined the total gravitating mass of the group, the hot gas mass, the
gas-to-stellar-mass ratio, and the baryonic fraction, if possible.  Four of
the groups in our survey (HCG 62, 68, and 97, and the NGC 2300 group) are at
least partially resolved with the PSPC, and so the masses obtained for those
groups are the most credible.  Three groups (HCG 12, 79, and 92) are
unresolved (also, HCG 79 is less than a 3$\sigma$ detection) and the
observation of HCG 94 is too far off axis to determine the spatial profile
of its emission.

	In this analysis, all point and slightly extended sources in the
PSPC field (stars, group galaxies, etc.) were masked off, and then the
center of the emission was determined by assuming that the emission was
symmetrically distributed.  As can be seen in Fig.~3, the extended emission
in these groups appears fairly symmetrical, which implies that the gas has
had time to relax into the group potential.  The elongation seen in the
X-ray isophotes of some groups (notably HCG 97) may be due to a somewhat
unrelaxed potential, but this level of disturbance will create only a
20\% uncertainty in the calculated mass, according to recent simulations
(Evrard, Metzler, \& Navarro 1995).

	The background level and maximum extent of the emission was
determined by examining the radial profile of the emission (see Figs. 5-9)
and the background level.  All of these groups have a radial extent of X-ray
emission of over 1\arcmin\ and none have a bright central point source.
Since the point-spread function of the PSPC at energies of 0.5-1.0 keV falls
to less than $10^{-3}$ of its central value for radii greater than
1\arcmin\ (Hasinger et al.~1992), the PSF does not affect our determination
of the background level or the radial extent of the emission.

	We fit the radial surface
brightness profile with a convolution of the point-spread function of the
PSPC with the standard hydrostatic-isothermal beta model (Cavaliere \&
Fusco-Femiano 1976):  $$S(r)=S_0\left( 1+\left( {r \over
r_{core}}\right)^2\right)^{-3\beta +0.5}$$
where $S_0$, $r_{core}$, and $\beta$ are free parameters.  This model, along
with the spectral fit for each group (see Table 5), allows one to solve for the
gas mass, and, again assuming hydrostatic equilibrium, for the gravitating
mass interior to the radius $r$:
$$M(r)=-{kT(r) r \over \mu m_p G}\left({d {\rm ln}(\rho_{gas})
\over d {\rm ln}(r)} + {d {\rm ln}(T) \over d {\rm ln}(r)}\right)$$
where the first derivative
equals $-3\beta$ and the second is assumed to be zero for all groups except
for HCG 62 (which is the only group bright enough for us to measure the
temperature gradient).  In galaxy clusters (and HCG 62), the temperature
gradient is small compared to $3\beta$, so our assumption of isothermality
does not have a large effect on the determination of $M(r)$.  The stellar
mass in galaxies was determined by assuming a mass-to-light ratio of
$M/L_B=8$ in solar units.  The mass-to-light ratio for galaxies in these
groups certainly will vary by Hubble type and interaction history (Rubin,
Hunter, \& Ford 1991; White et al.~1993, Worthey 1994), and the average
value for a group may be as low as $M/L_B=3$, but we will use a single
uniform ratio for the sake of simplicity and easy comparison.

	The determination of the gaseous mass of these systems is sensitive to
the abundance used, a quantity that usually cannot be obtained securely
from these data.  The reason that the gas mass depends upon the metallicity
is that line emission can be the main contributor to the X-ray spectrum at
$k$T $\sim$ 1 keV, so the inferred emission measure increases as one
adopts lower abundances.  For example, when fixing all other parameters in
a spectral fit, the emission measure rises by up to a factor of 2 as the
abundance is decreased from solar values to half solar.
As the mass is proportional to the square root of the emission measure,
uncertainties in the abundance can lead to inaccuracies in the gas mass
determination of up to 50\%.

	Another source of uncertainty in the gas mass estimate can be
traced to the background level used when fitting a beta model to the
system.  If the background level is taken to be less than the true
background, then the diffuse emission appears to extend further in radius
and has a shallower slope (lower value of beta).  The gaseous mass
estimate is increased by the larger radius, and to a lesser extent, by
the smaller beta value.  In extreme cases, this can lead to a factor of
two difference in the implied gas mass.  The group with the most uncertain
background level is HCG 62, which may have a background 10\% lower than the
level we chose.  The effect of this difference on our calculated quantities
is discussed extensively in \S 3.3.2.  The remaining groups have much
better determined backgrounds (see Fig. 5-9).

	The gravitational mass is considerably less sensitive to the
background level than is the gas mass.  If the background level is
underestimated and thus beta is smaller and the extent $r_{max}$ is
larger, the gravitational mass changes only slightly, since
$M_{grav} \sim \beta r_{max}$.  The choice of abundance does have
some effect on the gravitational mass since it is also proportional to
the temperature of the gas.  Lowering the abundance from solar to
half solar values will raise the fitted temperature, and thus $M_{grav}$,
by 10--20\%.

	For the groups with non-detections or very poor detections
of X-ray emission, we found upper limits on the amount of hot gas.  After
finding the net number N of photons within a circle of radius $r_g$ (the
radius of each group as given by Hickson [1982]) and the rms error $\sigma$,
we scaled the normalization factor of a 1 keV source
(fitted with an abundance of half the solar value)
in the same field by the ratio of N+3$\sigma$ (if N$>$0) or 3$\sigma$ (if
N$\leq$0) to the net counts of that 1 keV source.  From that scaled
normalization, and assuming that $\beta$=0.5, that the radial extent of the
emission is 0.2 Mpc, and that $r_{core}$ is 1/6 the radius of the
emission (along with the assumption that the temperature of the emission
is 1 keV, all conservative guesses given the results we found with the
well-detected diffuse emission), we calculated the upper limit
gas mass. The results of these analyses are listed in Table 6 and
discussed in greater detail below.

\subsubsection{HCG 12}

	The emission from this group is centered at the PSPC position
1$^h$27$^m$33\fs 0, -4\deg 40\arcmin 21\arcsec\ (throughout this
paper, coordinates are for epoch J2000.0), which is NNW of galaxy
{\it a}, halfway between galaxies {\it d} and {\it e}; and it has a radial
extent of 2\farcm 2 (0.19 Mpc) [Fig.~5].

	Even though the galaxies of this group are not resolved in this
observation, and thus the extended X-ray emission seen may be partially
or mostly emission from individual galaxies, we fit a beta model to it in
order to obtain an upper limit on the gaseous mass of the group and an
estimate of the total mass.  Using this fit as well as the spectral fit
above, we find that the total gravitating mass of this group is $1.4 \times
10^{13}$ M$_{\sun}$, with an upper limit of $2.1 \times 10^{13}$ M$_{\sun}$
if $\beta$=1.36---the $1\sigma$ upper limit (Table 5)---is used.  The mass
of hot gas is $2.2 \times 10^{11}$ M$_{\sun}$ = 1.6\% M$_{grav}$.  The
luminous mass of the galaxies in HCG 12 is $1.9 \times 10^{12}$ M$_{\sun}$ =
14\% M$_{grav}$, thus, the gas-to-stellar-mass ratio is 12\%\ and the total
baryon fraction is 15\%.  The uncertainties for these numbers are large
(note the uncertainty of $\beta$ in Table 5), mainly because the emission is
only moderately well-fit by a beta model.  This is certainly due to
unresolved galaxy emission being included in the fit.

\pagebreak

\subsubsection{HCG 62}

	We excluded the emission from the central galaxies (all emission
less than 3\arcmin\ from galaxy {\it a}---the center of the X-ray emission)
when examining the spatial and spectral properties of the diffuse emission
from this group.  We find a different spatial profile than did Ponman \&
Bertram (1993)[PB], obtaining a $\beta$ of $0.51 \pm 0.02$ rather than 0.36
and a core radius of 2\farcm 46 $\pm$ 0\farcm 15 rather than 0\farcm 5, and
tracing emission only out to a radius of 15\arcmin\ (0.36 Mpc) rather than
their value of at least 20\arcmin\ (Fig.~6).  Other differences are that PB
used an abundance of 0.3 times solar and a somewhat different fitting
procedure than we did (their fit to the radial profile of this group gave
$\beta=0.38$ and $r_{core}$=2\farcm 5; they then used a more elaborate model
to get the values above).  We find a radial temperature decrease consistant
with that of PB; $k$T=$1.24 \pm 0.14$ keV for an annulus extending over
radii of 3--8\arcmin\ to $1.07 \pm 0.08$ keV for an annulus at 8--15\arcmin.

	The total gravitating mass within a radius of 15\arcmin\ is $2.9
\times 10^{13}$ M$_{\sun}$, close to the value PB found at the same radius
($2.0 \times 10^{13}$ M$_{\sun}$).  However, we calculate a significantly
smaller gas mass, $8.1 \times 10^{11}$ M$_{\sun}$ within a radius of
15\arcmin\ (2.8\% M$_{grav}$), while they found $2.5 \times 10^{12}$
M$_{\sun}$ within a radius of 20\arcmin .  This difference may be
attributable to our choice of background level (producing a larger value of
$\beta$ and a smaller radial extent) and of a higher abundance than PB.
Decreasing the abundance towards zero significantly worsens our spectral
fit, but increases the gas mass up to 50\%.  The baryonic mass in galaxies
is the same in both our and PB's analyses: $7.3 \times 10^{11}$ M$_{\sun}$.

        Since the largest difference between our analysis and PB's is
the much smaller gas mass, we consequently found a
smaller baryon fraction and gas-to-stellar-mass ratio for HCG 62.
Our gas-to-stellar-mass ratio is 110\%, versus PB's value of
310\%, and our baryon fraction is 5.3\% at a radius of 15\arcmin , while
PB obtained a value of 11.7\% at the same radius.  This group has the largest
gas-to-stellar-mass ratio and the smallest baryon fraction of all the
groups in our sample.  In addition, our X-ray luminosity is less than that
stated by PB, but given our different radial extent and our exclusion of
the bright central 3\arcmin\, as well as PB's luminosity being bolometric
and ours being restricted to the PSPC band, this is not unexpected.

	We also reanalyzed these data for a background level 10\% below the
level originally chosen since this is a plausible level at large radii (see
Fig.~6a).  Reducing the background level increases the radial extent of the
emission to 27\arcmin\ (0.64 Mpc) and nearly doubles the net counts.  When
we fit a beta model to these data, the values of $\beta$ and $r_{core}$ did
not change substantially ($\beta = 0.48$, $r_{core}$=2\farcm 9), but the fit
was
significantly poorer than in our original analysis.  This new fit doubles
the gas mass and increases the total mass by nearly 60\%.  Since the stellar
mass remains the same, the gas-to-stellar-mass ratio becomes 210\% and
the baryon fraction falls to 5.1\%.

	Because our analysis leads us to believe that the background is
not well determined at large radii (\S 2), we believe that our original
background level of 0.080 counts/pixel is the most reasonable.  However,
for systems like HCG 62 with emission at large radii, additional off-axis
observations to determine the extent of the emission and the background
level would be of considerable value.

\subsubsection{HCG 68}

	The center of the diffuse emission is 1.5\arcmin\ southwest of galaxy
{\it a}.  The emission is elongated towards the northeast and southwest and
extends to a radius of 9\arcmin\ (0.13 Mpc) [Fig.~7].

	This group has the largest baryon fraction (19\%) but the lowest
measured gas-to-stellar-mass ratio of all the groups with detected diffuse
emission, which is the opposite extreme from HCG 62.

\subsubsection{HCG 79}

	Although this group is only detected at the $3\sigma$ level in its
PSPC observation, the method used to find upper limits on the gas mass in
groups was used to find a gas mass (see Table 6).  Consequently, the value
found is much more uncertain than those of the more X-ray luminous groups.
However, the gas-to-stellar-mass ratio of 5.5\%\ is in the same range as the
other groups.

\subsubsection{HCG 92}

	The center of the emission in HCG 92 is roughly halfway between
galaxies {\it b} and {\it c} (22$^h$ 36$^m$ 00\fs 1, +33\deg\ 58\arcmin\
23\arcsec ), and extends to a radius of 3\arcmin\ (0.11 Mpc).
Unfortunately, the emission of this group is very poorly fit by a beta model
(the best model had a reduced ${\chi}^2$ of 20.7) so we did not calculate
a gas or total gravitating mass for this system.  As is the case for HCG
12, a significant amount of the X-ray emission of this group is probably due
to unresolved emission from the galaxies rather than diffuse hot gas in the
potential well of the group.  Planned HRI observations of this group should
clarify the nature of the X-ray emission, and allow a determination of the
gas and total gravitational masses in this group.

\subsubsection{HCG 94}

	HCG 94 is an extremely bright X-ray source (L(0.07-3.0 keV) = $7.19
\times 10^{43}$ erg s$^{-1}$) with a high temperature ($k$T = $3.7 \pm 0.6$
keV) compared to other Hickson compact groups.  This group may be associated
with the nearby Abell cluster A2572 ($z$ = 0.0395), as noted by Rood \&
Struble (1994).  Abell 2572 is also seen in the HCG 93 PSPC image, although
it is even further off-axis than HCG 94.

\subsubsection{HCG 97}

	Extended emission is centered on galaxy {\it a} and extends
7\arcmin\ (0.27 Mpc) from it (Fig.~8).  It is difficult to separate the
emission from galaxy {\it a} from the diffuse emission, so we assumed that
the diffuse emission was dominant for radii greater than 48\arcsec\ (31
kpc).  This radius was chosen because it is twice the optical radius (at
$\mu_B$=25 magnitudes per square arcsecond [Hickson 1993]) of HCG 97a and it
is the radius outside of which the X-ray isophotes become more irregular.

\subsubsection{NGC 2300 group}

	We find that the diffuse emission in this group is centered
3\arcmin\ west of NGC 2300 and that it extends 17\arcmin\ (0.21 Mpc) from
this point (Fig.~9), rather than the 25\arcmin\ claimed by Mulchaey et al.
(1993)[MDMB].  The emission declines quite slowly with radius ($\beta$ =
0.32 $\pm$ 0.02), and the core radius is $2.19^{+0.35}_{-0.30}$ arcminutes
(the large uncertainty is due to the proximity of the emission from NGC 2300
to the center of the diffuse emission---a great deal of the diffuse emission
is masked out along with the galaxy's emission).

	We calculate a total gravitating mass for this group of $7.0 \times
10^{12}$ M$_{\sun}$, over a factor of 4 less than the mass found by MDMB
within a radius of 15\arcmin\ and a factor of 2.9 less than the total mass
they found within 25\arcmin\, reported in their note added in proof.  The
difference between their values and ours is likely due to differences in the
value of $\beta$, although this is unclear since MDMB did not publish a
value of $\beta$ or $r_{core}$.  MDMB assumed a uniform density of gas to
obtain an upper limit of $5 \times 10^{11}$ M$_{\sun}$ within a radius of
15\arcmin ; we fit a beta model with the parameters listed above and find
M$_{gas}$ = $5.2 \times 10^{10}$ M$_{\sun}$, or 0.7\%\ of the total mass.
With our standard mass-to-light ratio, the baryonic mass in the three
galaxies of the group is $9.2 \times 10^{11}$ M$_{\sun}$ = 13\% M$_{grav}$,
giving a gas-to-stellar mass ratio of 5.7\% and a total baryonic fraction
for this group of nearly 14\%, much greater than the 4\%\ value given in
MDMB.  (With the corrections made in their note added in proof, however,
their baryon fraction is 6\%).  The disparity between our results and those
of MDMB is probably related to our improved flattening and background
subtraction techniques.

\section{Further Discussion and Conclusions}

	Hickson Compact Groups are apparently quite poor in gas as compared
to richer groupings of galaxies.  While larger groups and clusters have gas
masses two to six times greater than the baryonic mass in their galaxies
(Arnaud et al.~1992, David et al.~1990), all but one of the groups we
analyzed have a gas-to-stellar-mass ratio of less than one-eighth, with a
median value of close to 5\%.  While this ratio is two orders of magnitude
less than that found in clusters, it is somewhat higher than the
0.5--4\%\ values found in bright early-type galaxies (Forman, Jones, \&
Tucker 1985).

        Even with this apparent shortage of gas relative to richer
structures, these compact groups have baryon fractions in the
range found in clusters and rich groups.  Baryon fractions for those
structures range from 10\%\ to over 30\%\ (Blumenthal et al.~1984; Briel,
Henry, \&\ B\"oringer 1992; David et al.~1994), while the fractions for the
four resolved groups in our study range from 5\%\ to 19\%.  The constancy of
the baryon fraction over all scales from individual galaxies to rich
clusters was noted by Blumenthal et al.~(1984), although more recent ROSAT
data seem to indicate that the baryon fraction may increase with increasing
mass scale (David, Forman, \& Jones 1994).

	If the errors on the parameters listed in Table 5 are taken into
account along with the possibility that $M/L_B$ could be as low as 3, the
gas-to-stellar-mass ratio for HCG 62, the system with the largest ratio,
could vary from 95\%\ to 340\%\, and that for HCG 68, the system with the
lowest measured ratio, could be from 1.6\%\ to 10\%.  The range of baryon
fractions for these two groups are then 4.9--6.4\%\ (HCG 62) and
5.8--24\%\ (HCG 68).  These are the largest possible errors we could
calculate from our data, but these two groups---which occupy the extremes of
baryon fraction and gas-to-stellar-mass ratio in our sample---are still
fairly distinct in these two quantities.  Note that the uncertainty in the
abundance is not taken into account above; changing the abundance by a factor
of 2 can change the gas mass by up to a factor of 50\%, as noted in \S 3.3.

It is important to keep in mind that comparisons
between these groups, rich clusters, and individual galaxies involves
properties measured at different density contrasts.  The
mean densities inferred from the binding mass estimates in Table 6
represent enhancements by factors $>\!2000$
over the mean background density (assuming $\Omega_0 \le 1$).
For rich clusters, the mean density within an Abell radius is roughly
an order of magnitude smaller, while the density contrast within a few
optical radii of a large elliptical can be an order of
magnitude larger.  It is not unreasonable to expect that properties
such as the gas-to-stellar-mass ratio will vary as a function of
radius from the group center or, equivalently, mean density contrast.

	One concern then is that the low gas-to-stellar-mass ratio may be
biased by the selection on optical compactness in Hickson's
catalogue.  More gas may lie well beyond the optical extent of the group.
For a value of $\beta$ = 0.5, characteristic of these groups, the gas mass
is proportional to $r^{1.5}$, while the stellar mass would increase linearly
with the radius if it possesses an isothermal distribution as do clusters.
At the absolute extreme, the detected galaxies could be the only galaxies in
the group, in which case the stellar mass is constant with radius.  Under
these assumptions, and if the group extends to 1 Mpc (representing a density
contrast $\sim \! 10^2$), the gas-to-stellar-mass ratio would increase by a
factor of 3--10.  Even with the largest increase, the median value of this
ratio would rise from about 5\%\ to 50\%, still an order of magnitude less
than that found in clusters of galaxies (David et al.~1990).  Since most of
the baryons are from stars in the group galaxies, changes in the baryon
fraction from this effect will be minor.

	Physical explanations for the low gas fractions
include very efficient star formation in these systems or expulsion of
ambient gas through galactic winds generated by bursts of star formation.
Accurate determination of the metallicity in the gas would be a
valuable diagnostic.  The gas-to-stellar-mass ratio could also be
increased if it were determined that some of the galaxies were
projected interlopers.  As discussed below, this is probably not a
large effect for the spiral-poor groups we find to possess
detectable, diffuse emission.

	Some of the groups in our sample may be superpositions of field or
loose poor group galaxies rather than bound entities.  While Hickson \& Rood
(1988) claim that only $\sim$1\% of HCG's should be chance alignments, Mamon
(1986) holds that roughly half are chance alignments within larger poor
groups and Hernquist, Katz, \& Weinberg (1994) suggest that HCG's may be
portions of galaxy filaments seen end-on.  An intermediate hypothesis
(Sulentic 1987; Diaferio, Geller, \& Ramella 1994) posits that they are
entities newly (and continuously) formed within poor groups and that
three-quarters are bound systems.  Of the five groups (38\% of our sample)
that have no evidence for extended X-ray emission, all contain galaxies with
morphological and/or kinematical peculiarities that strongly indicate
interactions (Mendes de Oliveira \& Hickson 1994).  This may only indicate
interactions between members of a loose group that are not bound to one
another, but it would be rash to declare all undetected groups unbound.
Since the emission measure is proportional to the square of the gas density,
a small decrease in the gas density would make a group undetectable with the
PSPC.  The eight groups that have extended emission are more likely to be
bound, especially those that are resolved into separate diffuse and galaxy
X-ray emission by the PSPC.  The two highly luminous groups HCG 62 and 94
may be evolving into the cores of poor clusters or groups.

	One piece of evidence that may discriminate between bound groups and
superpositions is the correlation we have found between extended X-ray
emission and low spiral fractions.  Five of the 13 groups in our sample have
spiral fractions of 50\% or greater, and only one of them (HCG 92) has a
possible detection of diffuse X-ray emission (note, however, that the
extended emission of HCG 92 was not well fit by a beta model, indicating
that emission from galaxies may be dominant).  Of the eight groups with
spiral fractions under 50\%, seven groups have extended emission and the
eighth, HCG 4, contains a bright AGN whose X-ray emission overwhelms any
possible diffuse emission.  This is similar to the results of Henry et
al.~(1995) on X-ray selected groups:  such groups have spiral fractions a
factor of two smaller than optically selected groups.  Hernquist, Katz, \&
Weinberg (1994) also suggest that spiral-rich HCG's may be projections,
while groups rich in early-type galaxies may be bound groups.  As more PSPC
and HRI observations of HCG's enter the archives, this negative correlation
between spiral fraction and diffuse X-ray emission can be tested more
thoroughly.

	HCG 62 appears to be a different object from the other groups in
this study.  (HCG 94 may be a similar group, but our observation of it is
spatially unresolved.  An on-axis PSPC observation of this group has been
performed and an HRI observation is planned; both should be enlightening.)
Not only is HCG 62 very luminous in X-rays with highly extended emission,
but it also has the highest gas-to-stellar-mass ratio by over a factor of 9
and the lowest baryon fraction by over a factor of 2.  As Ponman et
al.~(1994) noted, the newly discovered fossil galaxy group RX~J1340.6+4018
has X-ray properties very similar to those of HCG 62.  However, HCG 62 is
not exceptional in its optical properties---its two central early-type
galaxies overlap but do not show clear signs of interaction or merging
(Pildis \& Bregman 1995).  Thus, X-ray bright, gravitationally bound compact
groups may not be found easily in optical surveys, and X-ray surveys may
find as yet unrecognized bound galaxy groupings.

\acknowledgements

This material is based upon work supported under a National Science
Foundation Graduate Fellowship and NASA grant NAGW-2135. We thank S. Snowden
for the use of his CAST\_EXP program and his patient help with its
installation, and J. Turner and G. Rohrbach for their similar patience with
the installation of PROSCON.  We also acknowledge discussions with P.
Hickson, L. David, R. Mushotzky, J. Mulchaey, and D. Davis.  This research
has made use of the NASA/IPAC Extragalactic Database (NED) which is operated
by the Jet Propulsion Laboratory, Caltech, under contract with the National
Aeronautics and Space Administration, as well as The Einstein On-Line
Service, Smithsonian Astrophysical Observatory.

\clearpage

\begin{planotable}{lcccc}
\scriptsize
\tablewidth{460pt}
\tablecaption{Properties of groups analyzed}
\tablehead{
\colhead{Group} & \colhead{$z$} & \colhead{Distance (Mpc)} &
\colhead{N$_{\rm accordant}$} & \colhead{Comments}}
\tablecomments{Column 2 is the median redshift of each group and column 4
is the number of galaxies in each group with accordant redshifts.  Both
are from either Hickson (1993) or Mulchaey et al.~(1993).}
\startdata
HCG 2 & 0.0144 & 86.3 & 3 & our observation \nl
HCG 4 & 0.0280 & 168 & 3 & bright AGN in galaxy {\it a} \nl
HCG 10 & 0.0161 & 96.5 & 4 & our observation \nl
HCG 12 & 0.0485 & 291 & 5 & not well resolved, diffuse emission? \nl
HCG 44 & 0.0046 & 27.6 & 4 & --- \nl
HCG 62 & 0.0137 & 82.1 & 4 & diffuse emission (Ponman \& Bertram 1993) \nl
HCG 68 & 0.0080 & 48.0 & 5 & our observation, diffuse emission \nl
HCG 79 & 0.0145 & 86.9 & 4 & Seyfert's Sextet, unresolved \nl
HCG 92 & 0.0215 & 129 & 4 & Stephan's Quintet, unresolved \nl
HCG 93 & 0.0168 & 101 & 4 & our observation \nl
HCG 94 & 0.0417 & 250 & 7 & off axis, X-ray luminous \nl
HCG 97 & 0.0218 & 131 & 5 & diffuse emission \nl
N2300 group & 0.0070 & 42.0 & 3 & diffuse emission (Mulchaey et al.~1993) \nl
\end{planotable}

\clearpage

\begin{planotable}{lccc}
\tablewidth{450pt}
\tablecaption{X-ray observations of groups}
\tablehead{
\colhead{Group} & \colhead{Total Obs.~Time (ksec)} &
\colhead{Time Used (ksec)} & \colhead{N$_{\rm H}$}}
\tablecomments{Column 4 is the galactic neutral hydrogen column in the
direction of each group, in units of $10^{20}$ cm$^{-2}$.}
\startdata
HCG 2 & 20.17 & 20.17 & 4.08 \nl
HCG 4 & 9.33 & 9.33 & 1.55 \nl
HCG 10 & 16.10 & 16.10 & 5.18 \nl
HCG 12 & 10.79 & 10.79 & 4.05 \nl
HCG 44 & 4.67 & 4.67 & 2.13 \nl
HCG 62 & 19.72 & 19.52 & 2.87 \nl
HCG 68 & 14.99 & 14.99 & 1.03 \nl
HCG 79 & 5.29 & 4.91 & 3.96 \nl
HCG 92 & 20.88 & 13.41 & 7.53 \nl
HCG 93 & 16.76 & 16.27 & 5.36 \nl
HCG 94 & (16.76) & (16.27) & 5.10 \nl
HCG 97 & 13.89 & 13.89 & 3.62 \nl
N2300 group & 6.04 & 6.04 & 5.95 \nl
\end{planotable}

\clearpage

\begin{planotable}{lllcccc}
\scriptsize
\tablewidth{400pt}
\tablecaption{Group galaxies detected in X-rays and upper limits}
\tablehead{
\colhead{Galaxy} & \colhead{Name} & \colhead{Type} & \colhead{M$_B$} &
\colhead{$k$T (keV)} & \colhead{Net Counts} & \colhead{L$_X$}}
\tablecomments{Column 3 is the Hubble type of each galaxy from either
Hickson (1993) or Mulchaey et al.~(1993),
column 4 is its absolute blue magnitude as determined from the group distance
(Table 1) and its apparent blue magnitude in Hickson (1993), and column 7
is the fitted X-ray luminosity in units of $10^{40}$erg s$^{-1}$ for the
0.07-3.0 keV energy band.}
\startdata
HCG 2a & UCG 312 & SBd & -21.4 & 1 & $78 \pm 12$ & 2.4 \nl
HCG 2b & Mrk 552 & Sp? & -20.3 & 1 & $47 \pm 10$ & 1.8 \nl
HCG 2c & UCG 314 & SBc & -20.6 & 1 & $10.7 \pm 5.1$ & 0.32 \nl
HCG 4a & \nodata & Sc & -22.6 & AGN & $3153 \pm 64$ & $1.13 \times 10^3$ \nl
HCG 10a & NGC 536 & SBb & -22.3 & 1 & $64 \pm 16$ & 3.1 \nl
HCG 10b & NGC 529 & E1 & -22.2 & 0.5 & $59 \pm 12$ & 2.9 \nl
HCG 10c & NGC 531 & Sc & -20.8 & 1 & $<21.5$ & $<1.0$ \nl
HCG 10d & NGC 542 & Scd & -20.2 & 1 & $<10.6$ & $<0.51$ \nl
HCG 44a & NGC 3190 & Sa & -20.7 & 1 & $52 \pm 13$ & 0.61 \nl
HCG 44b & NGC 3193 & E2 & -20.6 & 0.5 & $72 \pm 16$ & 0.69 \nl
HCG 44c & NGC 3185 & SBc & -19.7 & 1 & $27 \pm 12$ & 0.34 \nl
HCG 44d & NGC 3187 & Sd & -19.1 & 1 & $22.4 \pm 6.5$ & 0.28 \nl
HCG 62a & NGC 4761 & E3 & -21.2 & $0.87 \pm 0.01$ & $5067 \pm 80$ &
199\tablenotemark{a} \nl
HCG 62b & NGC 4759 & S0 & -20.8 & \nodata & \nodata & \nodata \nl
HCG 62c & NGC 4764 & S0 & -19.0 & \nodata & \nodata & \nodata \nl
HCG 68a & NGC 5353 & S0 & -21.6 & $0.62 \pm 0.10$ & $717 \pm 29$ &
10.6\tablenotemark{b} \nl
HCG 68b & NGC 5354 & E2 & -21.2 & \nodata & \nodata & \nodata \nl
HCG 68c & NGC 5350 & SBbc & -21.5 & 1 & $119 \pm 15$ & 1.5 \nl
HCG 68d & NGC 5355 & E3 & -19.7 & 0.5 & $94 \pm 15$ & 0.75 \nl
HCG 68e & NGC 5358 & S0 & -19.2 & 0.5 & $<25.4$ & $<0.20$ \nl
HCG 93a & NGC 7550 & E1 & -22.4 & 0.5 & $41 \pm 15$ & 2.2 \nl
HCG 93b & NGC 7549 & SBd & -21.8 & 1 & $14.4 \pm 5.6$ & 1.2 \nl
HCG 93c & NGC 7547 & SBa & -21.1 & 1 & $15.9 \pm 5.3$ & 1.3 \nl
HCG 93d & \nodata & S0 & -19.7 & 0.5 & $<13.5$ & $<0.73$ \nl
HCG 97a & IC 5357 & E5 & -21.4 & $1.04 \pm 0.09$ & $395 \pm 22$ &
56.6\tablenotemark{c} \nl
HCG 97b & IC 5359 & Sc & -20.8 & 1 & $<66.5$ & $<9.5$ \nl
NGC 2300 & \nodata & E3 & -21.3 & $0.50 \pm 0.12$ & $301 \pm 19$ & 9.1 \nl
NGC 2276 & \nodata & Sc & -21.4 & $1.0 \pm 0.4$ & $98 \pm 12$ & 3.1 \nl
IC 455 & \nodata & S0 & -19.1 & 0.5 & $<15.7$ & $<0.47$ \nl
\tablenotetext{a}{The X-ray temperature and luminosity for HCG 62a, b, and
c are for a region 3\arcmin\ in radius centered on a---the individual
galaxies are not resolved.}

\tablenotetext{b}{HCG 68a and b are not resolved separately, so the X-ray
temperature and luminosity given are for both galaxies.}

\tablenotetext{c}{The X-ray temperature and luminosity for HCG 97a are for
a region centered on the galaxy and 48\arcsec\ in radius.}
\end{planotable}

\clearpage

\begin{planotable}{lccc}
\tablewidth{460pt}
\tablecaption{Basic properties of extended emission}
\tablehead{
\colhead{Group} & \colhead{Radial Extent} &
\colhead{Background Level} & \colhead{Net Counts}}
\tablecomments{Column 2 is the radial extent of the emission as discussed
in \S 3.3 and column 4 is the background level in counts per
(4\arcsec\ $\times$ 4\arcsec ) pixel.}
\startdata
HCG 12 & 2\farcm 2 (0.19 Mpc) & 0.014 & $96 \pm 23$ \nl
HCG 62 & 15\arcmin\ (0.36 Mpc) & 0.080 & $3603 \pm 318$ \nl
HCG 68 & 9\arcmin\ (0.13 Mpc) & 0.098 & $710 \pm 88$ \nl
HCG 79 & \nodata & 0.012 & $28.4 \pm 11.1$ \nl
HCG 92 & 3\farcm 0 (0.11 Mpc) & 0.040 & $501 \pm 56$ \nl
HCG 97 & 7\arcmin\ (0.27 Mpc) & 0.026 & $923 \pm 78$ \nl
N2300 group & 17\arcmin\ (0.21 Mpc) & 0.0086 & $838 \pm 131$ \nl
\end{planotable}

\clearpage

\begin{planotable}{lcccccc}
\scriptsize
\tablewidth{460pt}
\tablecaption{Fitted spectral and spatial properties of diffuse emission}
\tablehead{
\colhead{Group} & \colhead{$k$T (keV)} & \colhead{VEM} &
\colhead{$n_0$ (cm$^{-3}$)} & \colhead{$\beta$} &
\colhead{$r_{core}$ (arcmin)} & \colhead{L$_X$}}
\tablecomments{Column 3 is the volume emission measure divided by $4\pi$D$^2$
(where D is the distance to the group) in units of $10^{10}$ cm$^{-5}$,
column 4 is the central gas density, column 5 is the fitted value of
$\beta$, column 6 is the fitted value of the core radius, and column
7 is the fitted X-ray luminosity in units of $10^{41}$erg s$^{-1}$ for the
0.07-3.0 keV energy band.}
\startdata
HCG 12 & $0.72 \pm 0.36$ & $0.57 \pm 0.85$ & $5.7 \times 10^{-3}$ &
$ 0.92^{+0.44}_{-0.11}$ & $0.52^{+0.10}_{-0.20}$ & 8.27 \nl
HCG 62\tablenotemark{a} & ---\tablenotemark{b} & $43.2 \pm 2.8$ &
$1.5 \times 10^{-3}$ & $0.512 \pm 0.015$ & $2.46 \pm 0.15$ & 19.5 \nl
HCG 68 & $0.98 \pm 0.13$ & $3.4 \pm 1.3$  & $1.3 \times 10^{-3}$ &
${0.61 \pm 0.04}$ & $2.25^{+0.30}_{-0.20}$ & 1.21 \nl
HCG 97\tablenotemark{c} & $0.97 \pm 0.11$ & $5.9 \pm 1.5$ &
$6.0 \times 10^{-3}$ & $0.480^{+0.010}_{-0.015}$ & $0.22 \pm 0.05$ & 13.2 \nl
N2300 group & $0.93 \pm 0.20$ & $13.1 \pm 4.4$ & $2.8 \times 10^{-4}$ &
$0.321 \pm 0.015$ & $2.19^{+0.35}_{-0.30}$ & 2.92 \nl
\tablenotetext{a}{Excludes all emission within a radius of 3\arcmin\ of the
central galaxies.}
\tablenotetext{b}{This group has a measurable temperature gradient---see
the text.}
\tablenotetext{c}{Excludes all emission within a radius of 48\arcsec\ of the
central galaxy.}
\end{planotable}

\clearpage

\begin{planotable}{lccccc}
\scriptsize
\tablewidth{400pt}
\tablecaption{Component masses for compact groups}
\tablehead{
\colhead{Group} & \colhead{M$_{grav}$/M$_{\sun}$} &
\colhead{M$_{gas}$/M$_{\sun}$} & \colhead{Stellar Mass (M$_*$/M$_{\sun}$)} &
\colhead{M$_{gas}$/M$_*$} & \colhead{$f_{baryon}$}}
\tablecomments{Column 2 is the gravitating mass in solar masses, column 3
is the gas mass in solar masses, column 4 is the stellar mass in solar masses
assuming M/L$_B$=8, column 5 is the ratio of gas mass to stellar mass, and
column 6 is the baryon fraction (gas mass plus stellar mass, divided by the
gravitating mass).}
\startdata
HCG 2 & \nodata & $<3.8 \times 10^{10}$ & $8.3 \times 10^{11}$ &
$<4.6$\% & \nodata \nl
HCG 4\tablenotemark{a} & \nodata & \nodata & $1.8 \times 10^{12}$ & \nodata &
\nodata \nl
HCG 10 & \nodata & $<4.4 \times 10^{10}$ & $2.4 \times 10^{12}$ & $<1.8$\% &
\nodata \nl
HCG 12\tablenotemark{b} & $1.4 \times 10^{13}$ & $2.2 \times 10^{11}$ &
$1.9 \times 10^{12}$ & 12\% & 15\% \nl
HCG 44 & \nodata & $<2.2 \times 10^{10}$ & $6.0 \times 10^{11}$ & $<3.7$\% &
\nodata \nl
HCG 62 & $2.9 \times 10^{13}$ & $8.1 \times 10^{11}$ & $7.3 \times 10^{11}$ &
110\% & 5.3\% \nl
HCG 68 & $8.7 \times 10^{12}$ & $4.0 \times 10^{10}$ & $1.6 \times 10^{12}$ &
2.5\% & 19\% \nl
HCG 79 & \nodata & $3.5 \times 10^{10}$ & $6.3 \times 10^{11}$ & 5.5\% &
\nodata \nl
HCG 92 & \nodata & \nodata & $3.5 \times 10^{12}$ & \nodata & \nodata \nl
HCG 93 & \nodata & $<5.4 \times 10^{10}$ & $2.2 \times 10^{12}$ & $<2.5$\% &
\nodata \nl
HCG 97 & $1.4 \times 10^{13}$ & $1.5 \times 10^{11}$ & $1.5 \times 10^{12}$ &
10\% & 12\% \nl
N2300 group & $7.0 \times 10^{12}$ & $5.2 \times 10^{10}$ &
$9.2 \times 10^{11}$ & 5.7\% & 14\% \nl
\tablenotetext{a}{The AGN in HCG 4a is too bright to allow an
estimate of an upper limit on the gas mass.}
\tablenotetext{b}{M$_{gas}$ may be too large and M$_{grav}$ may not be
well-determined because of spatial confusion between galaxy emission
and the diffuse group emission.}
\end{planotable}

\begin{figure}

\caption{Background level as a function of distance from field center for
rp700112 (light line) and rp700375 (heavy line).  Each pixel is
4\arcsec\ on a side.  The dashed lines represent the best fit for the
average for 7\arcmin\ $<$ r $<$ 20\arcmin .}

\caption{Background level as a function of distance from field center
for two different flattening attempts on rp700117.  Each pixel is
4\arcsec\ on a side.  The light line
represents a full spectrum (0.1--2.2 keV) image with a soft energy flat
field, while the heavy line represents an image excluding energies
$<$ 0.5 keV with a hard flat field.  The dashed lines are best fits
for the average for 5\arcmin\ $<$ r $<$ 15\arcmin .}

\caption{Contour maps of smoothed (by a gaussian of FWHM=0.63\arcmin)
X-ray emission from compact groups.  Optical positions of group galaxies
are marked by asterisks and PSPC coordinates are
for epoch J2000.0. Contours are drawn at 1$\sigma$, 2$\sigma$, 4$\sigma$,
8$\sigma$, \ldots\ above the background level, where $\sigma$ is calculated
for the smoothed background. (a) HCG 12. (b) HCG 62. (c) HCG 68. (d) HCG 79.
(e) HCG 92. (f) HCG 97. (g) NGC 2300 group.}

\caption{Optical to X-ray luminosities for galaxies in compact groups.  Open
circles denote early-type galaxies with strong detections, filled circles
are early-type galaxies with weak detections; asterisks are late-type
galaxies with strong detections, and stars are late-type galaxies with weak
detections.  Error bars are from the error in the net counts, as listed in
Table 3.  Upper limits on the X-ray luminosity are denoted by symbols with
descending arrows.  The L$_X$-L$_B$ relation for early-type galaxies of
Bregman, Hogg, \& Roberts (1992), converted to the ROSAT passband, is shown
as a solid line; a linear relation appropriate for late-type galaxies is
shown as a dotted line.}




\caption{Radial profile of the extended emission in HCG 12. Each pixel
is 4\arcsec\ $\times$ 4\arcsec . (a) The
linear-linear plot shows the choice of background level (dashed line).
(b)  Radial profile of background-subtracted emission as log counts per
pixel versus log r.}

\caption{Radial profile of the extended emission in HCG 62.  See the caption
to Fig.~5. (a) The gap at $r\sim 20$\arcmin\ is due to the exclusion of the
support ring.  The dashed line is the background level we chose, while the
dotted line is a plausible background 10\% lower.  See the text for a fuller
explanation.}

\caption{Radial profile of the extended emission in HCG 68.  See the caption
to Fig.~5.}

\caption{Radial profile of the extended emission in HCG 97.  See the caption
to Fig.~5.}

\caption{Radial profile of the extended emission in the NGC 2300 group.  See
the caption to Fig.~5. (a) The gap at $r\sim 20$\arcmin\ is due to the
exclusion of the support ring.}

\end{figure}

\end{document}